# Revisiting the relationship of black-hole kicks and multipole asymmetries

Jannik Mielke,[*] Shrobana Ghosh, Angela Borchers, and Frank Ohme
*Max Planck Institute for Gravitational Physics (Albert Einstein Institute), D-30167 Hannover, Germany*
*and Leibniz Universität Hannover, D-30167 Hannover, Germany*

Precession in black-hole binaries is caused by a misalignment between the total spin and the orbital angular momentum. The gravitational-wave emission of such systems is anisotropic, which leads to an asymmetry in the $\pm m$ multipoles when decomposed into a spherical harmonic basis. This asymmetric emission can impart a kick to the merger remnant black hole as a consequence of linear momentum conservation. Despite the astrophysical importance of kicks, multipole asymmetries contribute very little to the overall signal strength and, therefore, the majority of current gravitational-wave models do not include them. Recent efforts have been made to include asymmetries in waveform models. However, those efforts focus on capturing finer features of precessing waveforms without making explicit considerations of remnant kick velocities. Here we close that gap and present a comprehensive analysis of the linear momentum flux expressed in terms of multipole asymmetries. As expected, large asymmetries are needed to achieve the largest kick velocities. Interestingly, the same large asymmetries may lead to negligible kick velocities if the antisymmetric and symmetric waveform parts are perpendicular to each other around merger. We also present a phenomenological tool for testing the performance of waveform models with multipole asymmetries. This tool helped us to fix an inconsistency in the phase definition of the IMRPhenomXO4a waveform model.

## I. INTRODUCTION

Anisotropic emission of gravitational waves (GWs) from coalescing binary black holes (BBHs) causes the remnant black hole (BH) to recoil. The GWs carry linear momentum away from the binary and as a consequence of the conservation of linear momentum, the remnant is kicked in a direction opposite to the GW linear momentum flux [1,2].

When the spins are not aligned with the orbital angular momentum, the GW emission can be highly anisotropic, producing kick magnitudes of the order of thousands of kilometers per second. In these systems, the in-plane spin components induce orbital and spin precession, which leads to modulations in the GW amplitude and phase [3,4].

The simplest binary configuration with maximum in-plane spin is the "superkick" configuration [5–8]. These are equal-mass binaries with the spin components purely in the orbital plane. In addition, the spins point in opposite directions. For superkick cases, the frame-dragging effect of the two antiparallel in-plane spins moves the center of mass up and down in an inertial frame [9]. This manifests itself in an asymmetry of the GW emission in the $+m$ and $-m$ multipoles when we decompose the GW strain into spin-weighted spherical harmonics. Analogously, we can also say that the waveform can be written as a sum of symmetric and antisymmetric parts with respect to a reflection on the orbital plane [10]. The kick is proportional to the multipole asymmetry [5]. It also strongly depends on the orbital phase when most of the linear momentum is radiated away, centered roughly around the merger time [5].

Some configurations lead to even higher kicks, and are called hang-up kick configurations [11–13]. These binaries have the same properties as superkick configurations, but the spins are slightly tilted above the orbital plane. Above (below) the orbital plane refers to spins that have positive (negative) components along the direction of the orbital angular momentum. Extrapolated to maximal spin, these configurations produce the largest kicks, reaching up to 5000 km/s. This is due to the combination of having high multipole asymmetry and the orbital hang-up effect [14]. The orbital hang-up effect postulates that equal-mass configurations with spins above the orbital plane emit a larger amount of linear momentum, angular momentum, and energy than configurations with spins below the orbital plane. The reason is, in the former case, a larger number of orbits are required until the merger is complete, compared to the latter case.

[*]Contact author: jannik.mielke@aei.mpg.de

Measurements of kicks and spins with GW observatories are of high astrophysical interest. Spin measurements could inform us about the origin of the observed BHs, as the misalignment of the spins could give insights into whether the BHs originate in isolated or dynamical formation channels [15–18]. Remnants with large kicks could be ejected from their host environments and thus, can influence the BBH merger rate [19,20].

The majority of observed signals have been insufficiently strong to permit the observation of either a kick or spin precession. The GW200129_065458 signal represents a notable exception [21,22], but unfortunately, it also exhibits some data quality issues [23,24]. Nevertheless, it is expected that the number of signals with meaningful spin and kick information will increase as the detector sensitivity improves. This is why the field of waveform modeling has already made significant advances in incorporating the related multipole asymmetries. Multipole asymmetry could possibly impact spin direction and magnitude measurements at high signal-to-noise ratios [25,26] or in combination with large kicks [27]. Additionally, a crucial understanding in waveform modeling is that the coprecessing waveform can be mapped into an equivalent aligned-spin waveform [28,29]. Neglecting multipole asymmetries is one of the clear limitations of the currently used mapping [30].

The time-domain waveform model NRSur7dq4 includes multipole asymmetries, as it relies on a basis interpolated directly from precessing numerical relativity (NR) waveforms [31], but is limited to mass ratio $q \lesssim 6$ and a total mass of $\lesssim 62 M_\odot$. The frequency-domain model of the dominant (2,2) multipole asymmetry [32], calibrated to a wider parameter space up to mass ratio $q = 8$ [33], is currently available only for IMRPhenomXO4a [34].

The objective of this study is to analyze the dependence of kicks on the interaction of the symmetric and antisymmetric parts of waveforms generated by precessing binaries. We investigate this relationship numerically using the NRSur7dq4 model, as it is the closest approximation to relativistic solutions of binaries in the strong field regime. In addition, we find a universal relationship that should be satisfied by waveform models that include multipole asymmetries. Future waveform models should incorporate the physics of the kick-multipole-asymmetry relation, which we found encoded in an analytical analysis and confirmed with NR simulations.

We start with a summary of previous works and adopt it to our notation in Sec. II. We discuss phenomenological observations of the kick-multipole-asymmetry relation and their usage for waveform modeling in Sec. III. In Sec. IV we analyze the most relevant parts of the linear momentum flux equation in terms of symmetric and antisymmetric waveform quantities.

## II. PRELIMINARIES

In this section we summarize the origin of the asymmetry and how it relates to out-of-plane kick. Throughout the text we use $G = c = 1$, but report kick velocities in metric units for ease of interpretation.

### A. Multipole asymmetries

To describe the GW emission of precessing binaries we have to solve the general relativistic two-body problem for Kerr BHs. On a quasicircular orbit the problem is characterized by seven intrinsic parameters: the mass ratio $q = m_1/m_2 \geq 1$ and the dimensionless spin vectors of the two BHs, $\vec{\chi}_1$ and $\vec{\chi}_2$.

Far away from the source, the gravitational field of the two-body problem is typically represented as a perturbation of the flat spacetime. When analyzed in the transverse traceless gauge, the independent components of this perturbation are commonly referred to as the $+$ and $\times$ polarizations of the GW emission. It is common to consider both polarizations simultaneously by defining a complex-valued scalar strain,

$$h = h_+ - i h_\times. \tag{1}$$

The strain can then be expanded in a basis of $s = -2$ spin-weighted spherical harmonics ${}^sY_{\ell,m}$ [35–37],

$$h(t, r, \theta, \varphi) = \sum_{\ell=2}^{\infty} \sum_{m=-\ell}^{\ell} {}^{-2}Y_{\ell,m}(\theta, \varphi) h_{\ell,m}(t, r), \tag{2}$$

where $(r, \theta, \phi)$ are usual spherical coordinates and $t$ denotes the time. The complex functions $h_{\ell,m}$ are called the multipoles of the GWs and depend on all the intrinsic source parameters.

The movement of binaries with spins parallel to the orbital angular momentum is confined to a fixed plane, which implies a reflection symmetry of the quadrupole moment over the orbital plane. Therefore, the GW multipoles satisfy

$$h_{\ell,-m} = (-1)^\ell h^*_{\ell,m}. \tag{3}$$

In the case of precessing binaries, the symmetry between the positive and negative $m$ multipoles is broken. Reference [10] showed that this multipole asymmetry is rotationally invariant. Hence the definition of the symmetric, abbreviated by the superscript “+,” and the antisymmetric, abbreviated by the superscript “–,” waveform multipoles,

$$h^\pm_{\ell,m} = \frac{1}{2}[h_{\ell,m} \pm (-1)^\ell h^*_{\ell,-m}], \tag{4}$$

is well defined in any inertial or rotating frame. An example of a rotating frame is the coprecessing frame, which tracks the direction that maximizes the GW emission in the dominant multipole, and simplifies the waveforms of precessing systems [10,38,39] by closely resembling that of a nonprecessing system.

We further introduce the $+/-$ amplitude, $|h^{\pm}_{\ell,m}| = a^{\pm}_{\ell,m}$, the $+/-$ phase, $\arg(h^{\pm}_{\ell,m}) = \phi^{\pm}_{\ell,m}$, and $+/-$ strain, $h^{\pm} = \sum {}^{-2}Y_{\ell,m} h^{\pm}_{\ell,m}$. Recalling the geometry of spin-weighted spherical harmonics, the symmetric waveform can be identified as the average of the GW radiation above and below the orbital plane, while the antisymmetric waveform is the respective difference.

Each multipole can be decomposed into a sum of its symmetric and antisymmetric part,

$$h_{\ell,m} = h^{+}_{\ell,m} + h^{-}_{\ell,m}. \tag{5}$$

By definition, the symmetric and antisymmetric multipoles satisfy convenient symmetry relations analogous to Eq. (3),

$$h^{+}_{\ell,-m} = (-1)^{\ell} h^{+^*}_{\ell,m}, \tag{6a}$$

$$h^{-}_{\ell,-m} = (-1)^{\ell+1} h^{-^*}_{\ell,m}, \tag{6b}$$

which makes it unnecessary to separately consider the negative $m$ symmetric/antisymmetric multipoles. Note that due to the linearity of the derivative operator, this symmetry also holds for $\dot{h}^{\pm}_{\ell,m}$, where the dot means the derivative with respect to time.

### B. Remnant kick velocity

When two BHs orbit each other, they emit GWs, which carry energy, angular and linear momentum away from the source. As they get closer, their orbital velocity increases, and in the final stages of their inspiral they merge into a single BH, releasing a large amount of GW radiation. As a consequence of momentum conservation, the remnant BH receives a kick in the opposite direction of the linear momentum flux, described in a local inertial frame as [36,40]

$$\frac{\mathrm{d}P_i}{\mathrm{d}t} = \lim_{r\to\infty} \frac{r^2}{16\pi} \oint \mathrm{d}\Omega l_i |\dot{h}|^2. \tag{7}$$

Here the standard solid angle element $\mathrm{d}\Omega$ and the unit radial vector in flat space

$$\vec{l} = (\sin\theta\cos\varphi, \sin\theta\sin\varphi, \cos\theta) \tag{8}$$

have been introduced. Using Eq. (5) for $\dot{h}$ and then evaluating the surface integral in Eq. (7) leads to an expression for the linear momentum flux in terms of the multipoles [41],

$$\frac{\mathrm{d}P_+}{\mathrm{d}t} = \lim_{r\to\infty} \frac{r^2}{8\pi} \sum_{\ell,m} \dot{h}_{\ell,m} (a_{\ell,m} \dot{h}^*_{\ell,m+1} + b_{\ell,-m} \dot{h}^*_{\ell-1,m+1} - b_{\ell+1,m+1} \dot{h}^*_{\ell+1,m+1}), \tag{9}$$

$$\frac{\mathrm{d}P_z}{\mathrm{d}t} = \lim_{r\to\infty} \frac{r^2}{16\pi} \sum_{\ell,m} \dot{h}_{\ell,m} (c_{\ell,m} \dot{h}^*_{\ell,m} + d_{\ell,m} \dot{h}^*_{\ell-1,m} + d_{\ell+1,m} \dot{h}^*_{\ell+1,m}). \tag{10}$$

Here $P_+ = P_x + iP_y$ has been introduced. The coefficients are given by

$$a_{\ell,m} = \frac{\sqrt{(\ell-m)(\ell+m+1)}}{\ell(\ell+1)}, \tag{11a}$$

$$b_{\ell,m} = \frac{1}{2\ell} \sqrt{\frac{(\ell-2)(\ell+2)(\ell+m)(\ell+m-1)}{(2\ell-1)(2\ell+1)}}, \tag{11b}$$

$$c_{\ell,m} = \frac{2m}{\ell(\ell+1)}, \tag{11c}$$

$$d_{\ell,m} = \frac{1}{\ell} \sqrt{\frac{(\ell-2)(\ell+2)(\ell-m)(\ell+m)}{(2\ell-1)(2\ell+1)}}. \tag{11d}$$

Integrating over the linear momentum flux and scaling with the remnant mass $m_f$ gives the final kick velocity

$$v_i = -\int_{-\infty}^{\infty} \mathrm{d}t \frac{\mathrm{d}P_i}{\mathrm{d}t} / m_f, \tag{12}$$

which is in the opposite direction of the total linear momentum flux.

Usually we choose the coordinate system so that, at some reference epoch, the $z$ axis is along the orbital angular momentum, the $x$ axis along the separation vector between the lighter and heavier BH, and the $y$ axis fulfills the triad. The in-plane kick refers to the component of the final kick velocity that is within the orbital plane of the BBH system just before merger. Besides the mass ratio, the in-plane kick is influenced by the spin components that are aligned or antialigned with the orbital angular momentum. Thus, in-plane kicks occur even for nonprecessing BBHs [42].

On the other hand, the out-of-plane kick refers to the component of the final kick velocity that is perpendicular to the orbital plane. The asymmetric emission between the positive and negative $m$ multipoles in precessing systems, influenced by the spin components not aligned with the orbital angular momentum, leads to the emission of linear momentum in the $z$ direction. In other words, the out-of-plane kick is related to the $+/-$ components of the waveform.

Boyle *et al.* [10] and Ma *et al.* [43] initiated an analysis of this relationship. Using the decomposition of the strain into symmetric and antisymmetric parts, we get

$$|\dot{h}|^2 = |\dot{h}^+ + \dot{h}^-|^2 = |\dot{h}^+|^2 + 2\mathrm{Re}(\dot{h}^- \dot{h}^{+*}) + |\dot{h}^-|^2. \quad (13)$$

Substituting Eq. (13) into Eq. (7), and noting that the nonmixed terms vanish, $\oint \mathrm{d}\Omega l_z |\dot{h}^+|^2 = \oint \mathrm{d}\Omega l_z |\dot{h}^-|^2 = 0$, gives

$$\frac{\mathrm{d}P_z}{\mathrm{d}t} = \lim_{r\to\infty} \frac{r^2}{8\pi} \oint \mathrm{d}\Omega l_z \mathrm{Re}(\dot{h}^- \dot{h}^{+*}). \quad (14)$$

That is Eq. (34) in Boyle *et al.* adapted to our notation. It is equivalent to Eq. (16) in Ma *et al.*, when only the dominant multipole asymmetry is included.

Ma *et al.* further analyzed equal-mass and equal-spin magnitude BBH systems with antiparallel spins, i.e., superkick configurations when the spins are in the orbital plane. They find that the magnitude of the out-of-plane kick is determined by the phase difference between the mass and the current quadrupole wave. This is equivalent to the phase between the $+/-$ parts of the waveform for the dominant multipoles, except for some prefactors. In the post-Newtonian (PN) regime, the phase difference is proportional to the azimuthal angle $\varphi$ of one of the spins. This is consistent with the sinusoidal dependence of the kick on $\varphi$ close to merger in superkick configurations [5–8].

In Sec. IV we present further results that build on their findings and analyze in more detail which interactions between the symmetric and antisymmetric parts of the waveform are responsible for the out-of-plane kick for arbitrary precessing systems.

## III. PHENOMENOLOGICAL OBSERVATIONS OF THE KICK MULTIPOLE-ASYMMETRY RELATION

In this section, we explore the morphology of the correlation between the out-of-plane kick and the dominant multipole asymmetry. We relate this morphology to intrinsic parameters of the BBH system and show how it can be used for testing the performance of waveform asymmetry models.

### A. Effect of the spin direction

To show the strong dependence of the kick on the spin orientation, we compute NRSur7dq4 waveforms for a set of $2^{16}$ equal-mass binaries where, for simplicity, only one BH has spin. The spin magnitude is fixed at $|\vec{\chi}| = 0.8$ and the spin direction is sampled over the spin tilt $\theta \in [0, \pi]$ and the spin azimuthal angle $\varphi \in [0, 2\pi]$, as illustrated in Fig. 1. The spins are defined close to merger at a reference frequency of $Mf_{\mathrm{ref}} = Mf_{\mathrm{ISCO}} = 1/(6^{3/2}\pi)$, where $M = m_1 + m_2$ is the total mass of the binary.

In Fig. 2, the results are visualized by a scatter plot of the kick velocity $v_z$ versus the maximum value of the dominant antisymmetric amplitude in the coprecessing frame

$$\hat{a}^- = \max_t [a^{-,\mathrm{copr}}_{2,2}(t)], \quad (15)$$

which we will sometimes refer to as "asymmetry" for simplicity. In some instances, these types of plots are designated as "spin-asymmetry-kick" plots.

The kick is computed with the SCRI [44] implementation of Eq. (10) by inserting all available multipoles in the coprecessing frame. Note that this means that the $z$ component of the kick refers to the direction of maximum emission, which varies with time. The frame transformation is also performed with SCRI.

Each point in Fig. 2 corresponds to one configuration from the sample set. The colors of the points indicate the in-plane spin component, i.e. the component of the spin perpendicular to the orbital angular momentum $\chi_\perp = \sin(\theta) \cdot \chi$. The linear dependence of the asymmetry on $\chi_\perp$ is clearly visible by the color gradient from left to right. This behavior is expected from PN and continues until merger [32].

The intuitive assumption that the kick vanishes when the asymmetry is switched off can be confirmed. As the asymmetry increases, the kick range becomes wider. A large asymmetry is not sufficient to generate large kick velocities. The remaining parameter, the in-plane spin angle $\varphi$, causes the large fluctuations in the kick profile for configurations with the same asymmetry. This is consistent with Pretorius' explanation that the kick depends strongly on the in-plane spin orientation at the time of coalescence due to frame-dragging addition or subtraction [9].

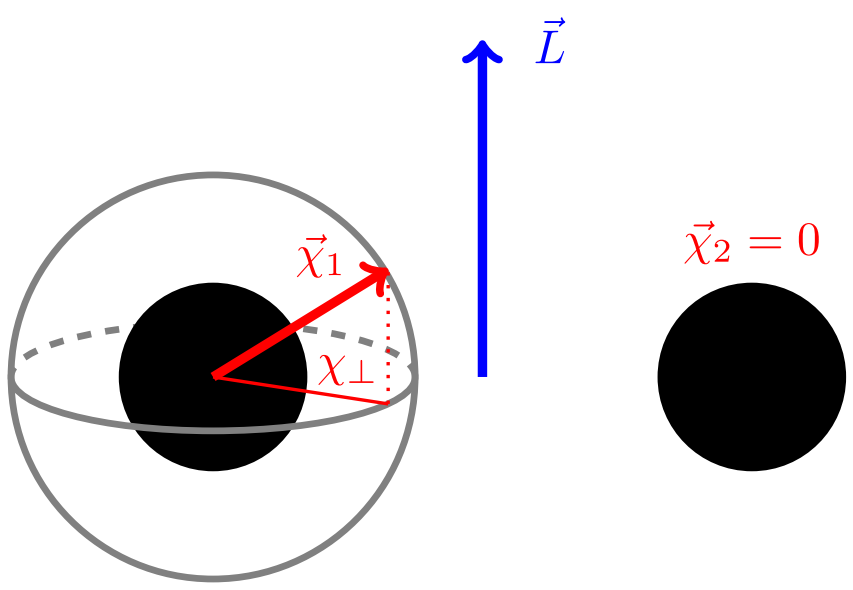

FIG. 1. Illustration of the parameter space sampling to generate waveforms for Figs. 2 and 4. The spin magnitude of one BH is fixed and the spin direction is sampled over the whole sphere. The second BH is not spinning. Spins are defined at the ISCO frequency. Additionally, in Fig. 3, we vary the mass ratio and place a spin on the lighter black hole.

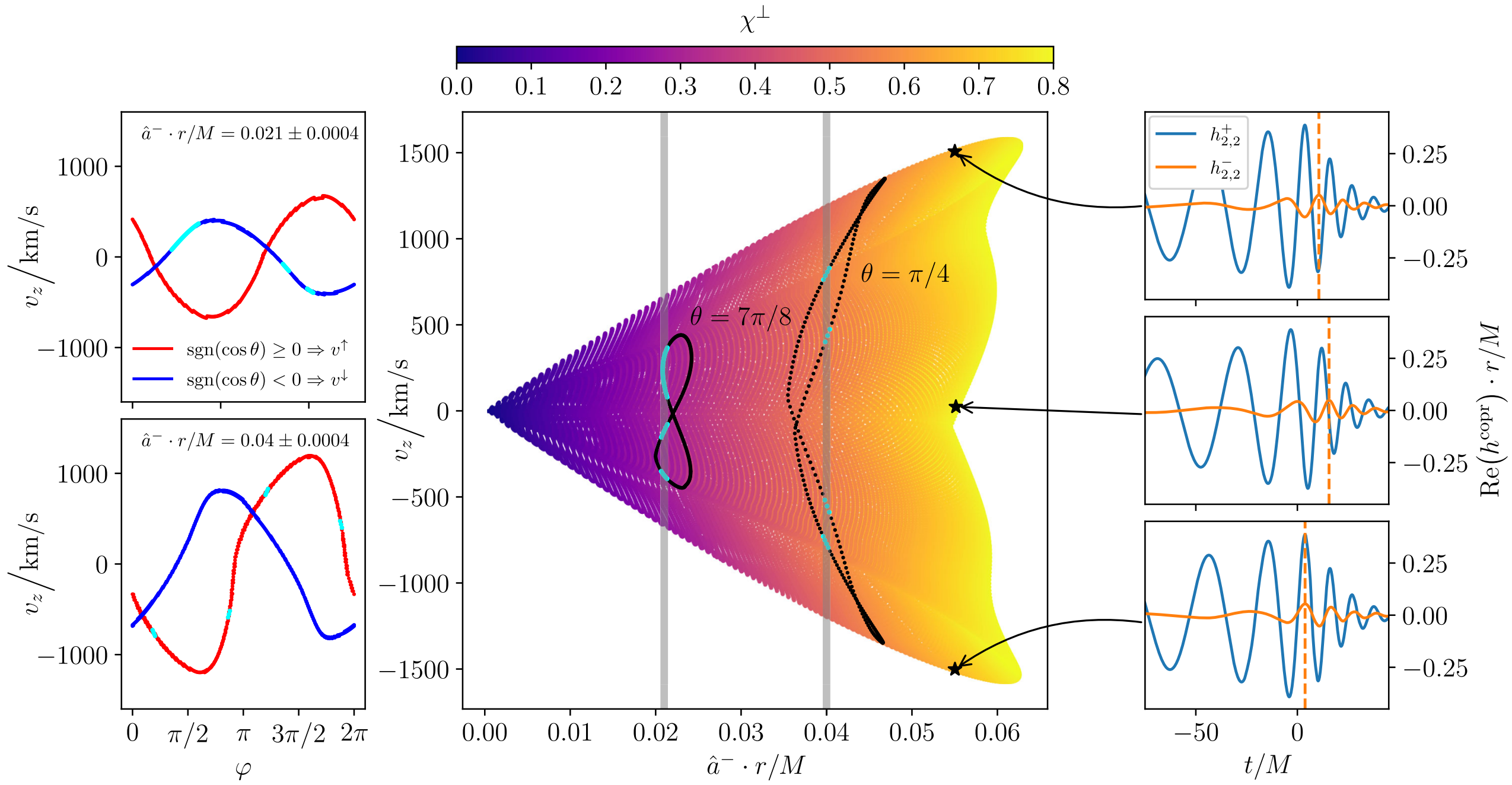

FIG. 2. Spin-asymmetry-kick correlation calculated from a large set of NRSur7dq4 waveforms from equal-mass binaries with varying spin directions. The central panel shows the kick along the axis of the largest GW emission calculated by all multipoles in the coprecessing frame versus the peak of the coprecessing antisymmetric waveform. The color map indicates the in-plane spin magnitude $\chi_{\perp}$. The black loops show $\theta = \pi/4$ and $\theta = 7\pi/8$ as an example for a constant value of the out-of-plane angle. The left panel shows the sinusoidal dependence of the kick on $\varphi$ for two slices of constant asymmetry. The red lines correspond to BBH configurations with the spin above the orbital plane, while the blue lines correspond to spins below the orbital plane. Cyan dots highlight the intersection of the chosen constant asymmetries with constant $\theta$ values. The real part of the dominant symmetric and antisymmetric waveform is shown in the right panel for three configurations with a large asymmetry and only a varying correlation between the symmetric and antisymmetric phases, resulting in a different kick velocity.

The $\varphi$ dependence is also illustrated for constant $\hat{a}^{-} \cdot r/M = 0.021$, 0.04 in the left panel of Fig. 2. These two slices with constant asymmetries have a small width of less than 2% in order to intersect with enough points, since we are dealing with a discretized space. The intersecting points then have two free parameters left to analyze, the in-plane spin angle and the kick velocity. The kick velocity has a sinusoidal dependence on the in-plane angle [8,11–13,42]. There are two sinusoids for configurations with the spin above, $v^{\uparrow}(\varphi)$, or below the orbital plane, $v^{\downarrow}(\varphi)$, characterized by the sign of $\cos\theta$. The peak amplitude of both $v^{\uparrow}$ and $v^{\downarrow}$ increases with the asymmetry. The fact that there are two sinusoids means that for a constant asymmetry there are always two out-of-plane angles, since $\hat{a}^{-} \sim \sin(\theta)$ and the sin function is symmetric around $\theta = \pi/2$. The peak amplitude of $v^{\uparrow}$ is larger than $v^{\downarrow}$. This can be explained with the orbital hang-up effect [14]. A system with spins above the orbital plane radiates more linear momentum compared to the corresponding mirrored system with spins below the orbital plane, since the merger of the BHs is delayed or prompted, respectively.

The small dependence of $\hat{a}^{-}$ on the in-plane spin angle explains the curved right edge of the spin-asymmetry-kick correlation plot. By highlighting loops of constant out-of-plane angle for $\theta = \pi/4$ and $\theta = 7\pi/8$ we can further observe that this behavior runs through the whole morphology. Each point on a black ring corresponds to a different in-plane spin angle. Spins above the orbital plane have a kick variation over the full range. Spins below the orbital plane cover a small range. This behavior is better seen in Fig. 3. Additionally, the intersections of the points on the loops of constant $\theta$ with the constant lines of asymmetry are plotted in cyan. They are also shown on $v^{\uparrow}$ respectively $v^{\downarrow}$.

We select three points with the same large asymmetry of $\hat{a}^{-} \cdot r/M \approx 0.055$ and plot their $+/-$ waveforms on the right panel of Fig. 2. The orange dashed line indicates the time of the antisymmetric peak amplitude and is slightly after $t = 0$, the time when the total NRSur7dq4 waveforms amplitude peak occurs. We observe that for very large positive or negative kicks there is no phase difference modulo $\pi$ between $\phi^{-}_{22}$ and $\phi^{+}_{22}$ in the merger-ringdown regime, while for the zero kick case there is a phase difference of about $\pi/2$. This is consistent with the results found in Ref. [43], but from a different phenomenological kick-multipole-asymmetry perspective. We will analyze the phase difference in more detail in Sec. IV C.

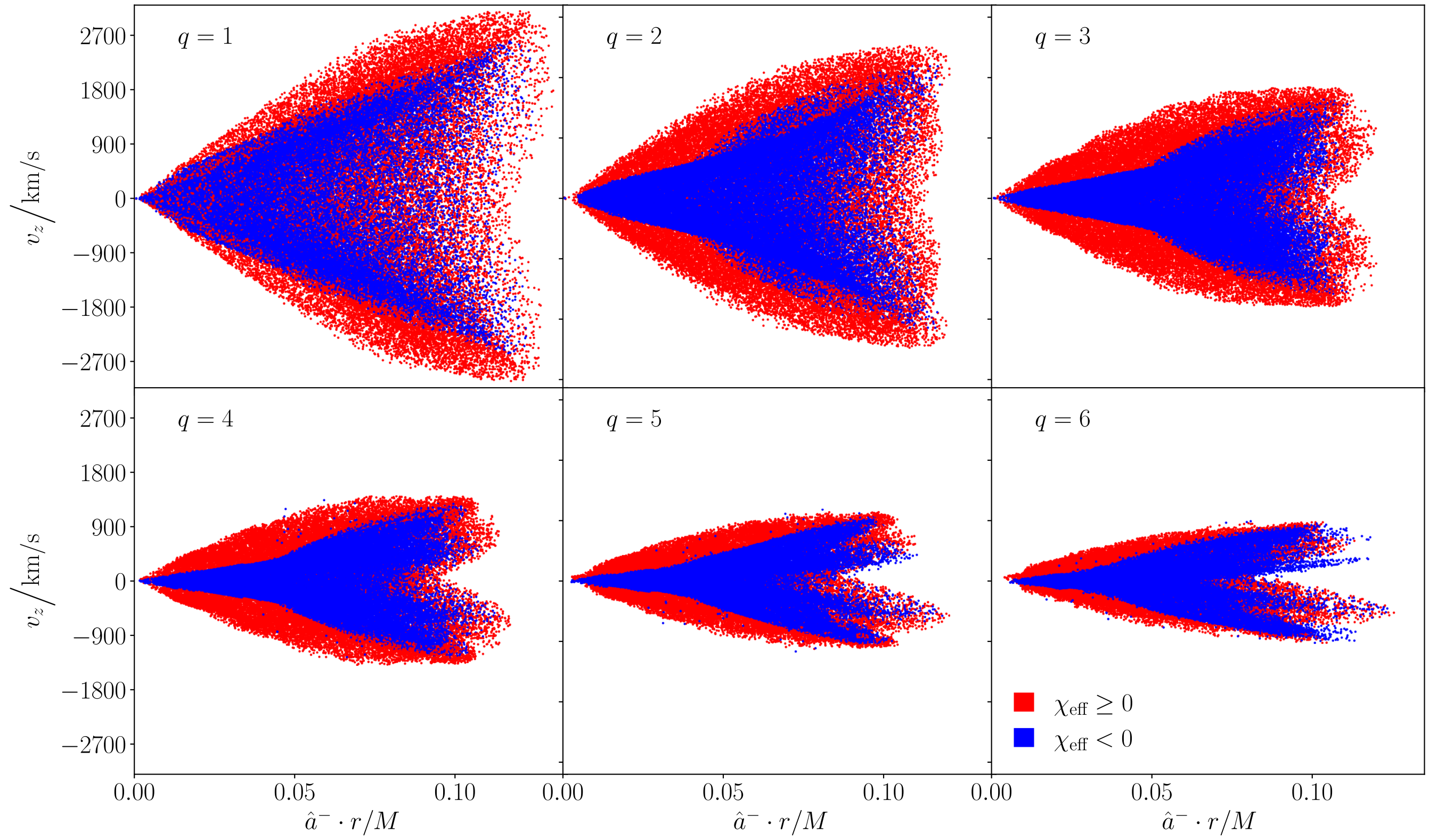

FIG. 3. Spin-asymmetry-kick correlation for different mass ratios calculated with NRSur7dq4. Spins of constant magnitude are placed on both black holes. Both spin directions are sampled over the entire sphere. The color indicates whether $\chi_{\mathrm{eff}}$ is positive (red) or negative (blue).

Several properties do not affect the morphology of the spin-asymmetry-kick correlation plot. First, the reference time or frequency at which we define the spins has no influence, since we sample over all spin directions. Nevertheless, choosing a reference frequency close to the merger leads to a clearer distribution of the in-plane spin angle within the envelope of the overall morphology. In addition, the kick is highly dependent on the exact orbital phase at coalescence of the two black holes [9]. Motivated by the fact that measuring spins near the merger leads to an improvement in the parameter recovery of the in-plane spin components [45], we decided to choose the Schwarzschild innermost stable circular orbit (ISCO) frequency as the reference frequency, $Mf_{\mathrm{ISCO}} = 1/(6^{3/2}\pi)$.

Second, the spin magnitude $\chi$ only changes the scale of the axes, since the out-of-plane kick and dominant antisymmetric amplitude is proportional to $\chi$. That is at least true in the PN regime [32,42], but within a good approximation the trend continues to the merger regime.

Lastly, we found that the ratio of the antisymmetric and symmetric peaks as shown in Fig. 1 of Ref. [27] does not change the morphology in Fig. 2. This is due to the definition of the symmetric waveform, where we explicitly want to put all the in-plane spin dependence in the antisymmetric waveform. Since we are fixing the mass ratio and the spin magnitude, the very small fluctuations of the symmetric amplitude peak from the spin direction have only a minor effect.

### B. Effect of the mass ratio

We have already studied and made statements about the influence of spin on the kick-multipole-asymmetry correlation. To cover the full intrinsic seven-dimensional parameter space, we reproduce the central plot of Fig. 2 but with different mass ratios and spins placed on both black holes. Both spin magnitudes are again fixed at $|\vec{\chi}_i| = 0.8$. The spin directions are sampled over the entire sphere.

The color map now indicates if the effective spin [46],

$$\chi_{\mathrm{eff}} = \frac{q\chi_1^z + \chi_2^z}{1 + q}, \tag{16}$$

is positive or negative. So it indicates whether the mass-ratio weighted total spin is above or below the orbital plane. As can be observed in Fig. 2, the black loops exhibit a discernible difference in kick range between spins above and below the orbital plane, which agrees with the orbital hang-up effect. Here in Fig. 3, however, the color distribution makes this fact more obvious.

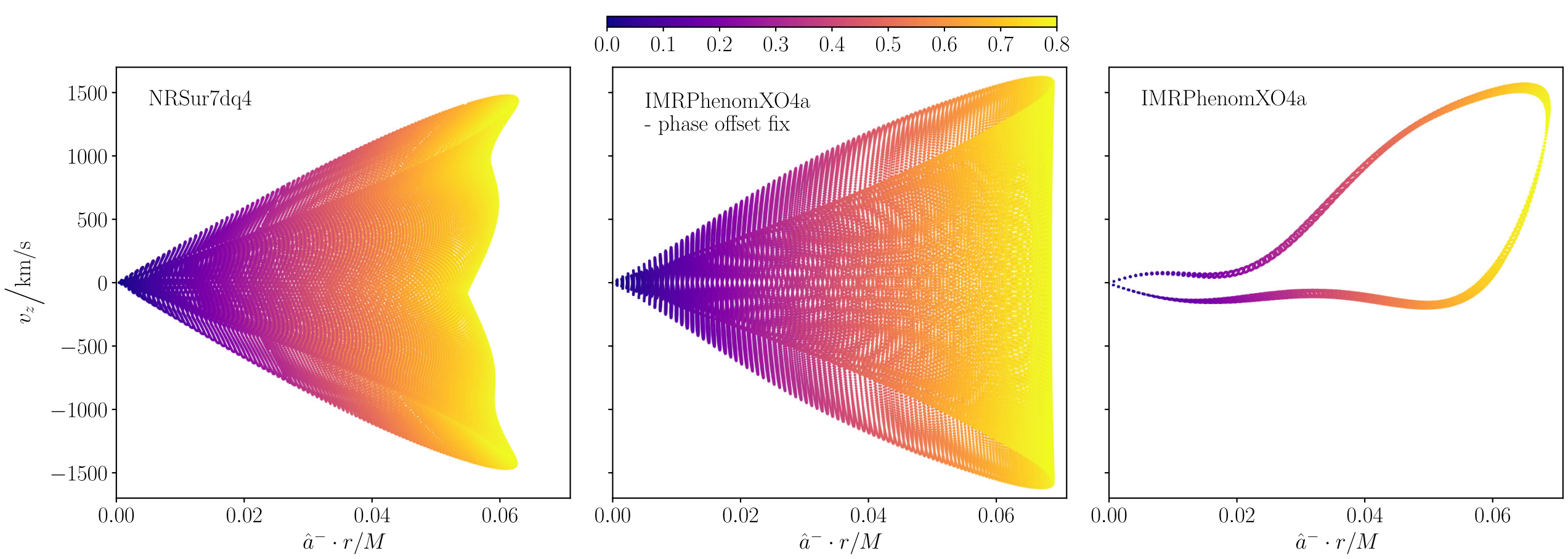

FIG. 4. Spin-asymmetry-kick correlation calculated by inserting only the $(2, \pm 2)$ multipoles in the coprecessing frame into Eq. (10). Different waveform models have been used: NRSur7dq4 (left), IMRPhenomXO4a with the phase offset fix described in the text (middle), and the IMRPhenomXO4a version currently implemented in LALSimulation (right).

For $q = 1$, varying the spin direction of two black holes only results in different scales of the axes. A larger kick and a larger asymmetry can be obtained. The influence of the $+/-$ phase difference can also be observed by choosing configurations with large asymmetry and different kick velocities. In addition, by looking at the spin sum, analogous statements can be made about the distribution of points with respect to their spin direction.

A stronger effect on the morphology is caused by the mass ratio. In PN, the asymmetry amplitude depends on $q$ [cf. Eq. (14) of Ref. [32] ]

$$a_{2,2}^{-,\mathrm{copr}} \sim \frac{2q^2}{(1+q)^3} \tag{17}$$

and reaches its maximum at $q = 2$. Since we choose a value of asymmetry in the merger-ringdown regime, we can no longer trust PN. However, we can at least see the trend of decreasing asymmetry in the calibrated region of the NRSur7dq4 model up to $q = 4$.

The kick dependence on $q$ changes the morphology more significantly, in the sense that the maximum achievable kick decreases with $q$. Compare this behavior for example with the PN motivated fitting formula in Eq. (22) of Ref. [47]. Another approach is to use the results below in Sec. IVA and identify the maximum achievable kick amplitude $\beta$, defined in Eq. (18), as the key factor for this behavior. $\beta$ decreases with increasing mass ratio because the symmetric amplitude $a^{+}_{\ell,m}$ decreases with increasing mass ratio.

For $q = 1$ system hang-up kick configurations produce the largest kicks with values above 3000 km/s. Interestingly, the asymmetry for these configurations does not reach the global maximum. Hang-up kicks combine the in-plane spin effect, which maximizes the antisymmetric waveform, and the orbital hang-up effect, which maximizes the duration of the GW emission due to the partial alignment of the spin with the orbital angular momentum.

Note two possibly nonphysical artifacts visible in Fig. 3 due to a choice of extreme parameters at the boundary or outside the calibration range of NRSur7dq4. First, for $q \geq 4$ we see blue points outside the envelope. Second, $\hat{a}^-$ and $v_z$ should be exactly zero for aligned-spin systems independent of the mass ratio. Such systems should always be at the origin of the plot, which is obviously not the case.

Overall, we see that the spin-asymmetry-kick correlation is a universal feature in the entire parameter space that can be used to improve and test waveform models.

### C. IMRPhenomXO4a: Case study for asymmetry model

Since the out-of-plane kick depends entirely on the multipole asymmetries, the spin-asymmetry-kick correlation plot in the middle panel of Fig. 2 is an effective tool to check the accuracy of the asymmetry model in IMRPhenomXO4a. Accordingly, in Fig. 4, we have reproduced the spin-asymmetry-kick plot with precisely the same parameters, but we have used IMRPhenomXO4a to generate the waveforms. Additionally, only the dominant multipoles are included in the kick calculation, as IMRPhenomXO4a does not incorporate subdominant multipole asymmetries. The inverse Fourier transform is performed on each multipole individually, after which the $+/-$ waveform definition is applied.

Prior to utilizing Fig. 4 to formulate conclusions regarding the IMRPhenomXO4a asymmetry model, it is advisable to undertake a brief comparison between the NRSur7dq4 plot displayed in the right panel of Fig. 4 and the NRSur7dq4 plot presented in Fig. 2. The data indicate a difference of approximately 150 km/s in the maximal reachable kick. This discrepancy is due to the exclusion of higher multipole asymmetry, which underscores the necessity for modeling all multipole asymmetries to fully describe the out-of-plane kick.

The right panel of Fig. 4 shows a very different morphology for IMRPhenomXO4a than that seen with NRSur7dq4. Notably, the whole range of kick velocities due to the rotation of the spin vector in the plane for a fixed in-plane spin magnitude, denoted by the black dotted loops in Fig. 2, is not reproduced by IMRPhenomXO4a. This demonstrates that the asymmetry model in IMRPhenomXO4a is not able to capture the sensitivity of the kick to in-plane spin rotation, as described before in Sec. III A. Since the amplitude of the asymmetry changes very little—if at all—due to in-plane spin rotation, we inspect the phase of the antisymmetric waveform more closely.

The phase of the antisymmetric waveform is modulated by the evolution of the spin vector in the orbital plane. The precession angle $\alpha$ tracks this evolution and the angle subtended by the in-plane spin vector with the radial vector is crucial in determining the value of $\alpha$ at some reference frequency of the waveform. Therefore, the phase coefficient $\phi_{A0}$ of the antisymmetric waveform in IMRPhenomXO4a [cf. Eq. (24) on [32] ] must be set to the correct value of $\alpha$ at the reference frequency of the waveform.

All PhenomX precessing approximants generate waveforms in the inertial $\mathbf{J}$ frame. In this frame the $z$ axis coincides with the direction of the total angular momentum $\mathbf{J}$ of the binary, which remains constant to a good approximation. The inertial waveform is produced by twisting up the multipoles calculated in the noninertial $\mathbf{L}$ frame, with the precession angles. The $\mathbf{L}$ frame tracks the precession of the waveform and coincides momentarily with the inertial $\mathbf{L}_0$ frame, which points to the initial angular momentum of the binary at a chosen reference frequency. Therefore, for the precession angle $\alpha$ to be initialized with the correct offset, it must (1) include the rotation of the $\mathbf{L}_0$ frame into the $\mathbf{J}$ frame at the reference frequency and (2) the rotation by angle $\zeta$ [cf. Eq. (C22) in Appendix C [48] ] that aligns the $x$ axis of the $\mathbf{J}$ frame with the separation vector at the reference frequency. Upon close inspection of the phase implementation of the antisymmetric waveform in IMRPhenomXO4a in LALSuite, we found the second piece missing. Figure 4 shows the impact of adding this piece to the asymmetry model in IMRPhenomXO4a and also shows that spin-asymmetry-kick correlation of IMRPhenomXO4a and NRSur7dq4 are in excellent agreement.

Note that the curved right edge of the NRSur7dq4 plots is due to a dependence of the antisymmetric amplitude on the in-plane spin $\varphi$. The asymmetry model used in IMRPhenomXO4a accounts for in-plane spin rotation by a fixed offset to the phase and therefore does not show the variation in the asymmetry amplitude. Investigations based on a family of NR simulations, sampling the in-plane spin angle at the same starting frequency, would be informative and will be studied in future work. We note here the efficacy of the spin-asymmetry-kick correlation plot as tool to test the nuances of the antisymmetric component of the waveform that are missed by typical performance markers used in waveform modeling, such as mismatches and parameter estimation tests.

## IV. INTERACTION OF THE SYMMETRIC AND ANTISYMMETRIC WAVEFORM

In the last section we explored phenomenologically how the BH spins affect the kick-multipole-asymmetry relation. We now want to study this relation more quantitatively. In particular, we focus on the interactions of different quantities of the $+/-$ waveform.

### A. Out-of-plane kick in terms of multipole asymmetries

We introduce a short-hand notation for combinations of different quantities of symmetric and antisymmetric parts of the waveform, which we will need in the following for indices $m^- = m^+$ and $|\ell^- - \ell^+| \leq 1$:

$$\alpha_{\ell^-,m^-,\ell^+,m^+} = \dot{a}^-_{\ell^-,m^-}\dot{a}^+_{\ell^+,m^+}, \tag{18a}$$

$$\beta_{\ell^-,m^-,\ell^+,m^+} = a^-_{\ell^-,m^-}a^+_{\ell^+,m^+}\dot{\phi}^-_{\ell^-,m^-}\dot{\phi}^+_{\ell^+,m^+}, \tag{18b}$$

$$\gamma_{\ell^-,m^-,\ell^+,m^+} = -a^-_{\ell^-,m^-}\dot{a}^+_{\ell^+,m^+}\dot{\phi}^-_{\ell^-,m^-}, \tag{18c}$$

$$\delta_{\ell^-,m^-,\ell^+,m^+} = \dot{a}^-_{\ell^-,m^-}a^+_{\ell^+,m^+}\dot{\phi}^+_{\ell^+,m^+}, \tag{18d}$$

$$\psi_{\ell^-,m^-,\ell^+,m^+} = \phi^-_{\ell^-,m^-} - \phi^+_{\ell^+,m^+}. \tag{18e}$$

It will turn out that among these combinations the possible achievable kick amplitude $\beta_{\ell^-,m^-,\ell^+,m^+}$ and the phase difference $\psi_{\ell^-,m^-,\ell^+,m^+}$ will be of particular importance. The short-hand notation allows us to write

$$\begin{aligned}
&\mathrm{Re}(\dot{h}^-_{\ell^-,m^-}\dot{h}^{+^*}_{\ell^+,m^+}) \\
&= \mathrm{Re}(\dot{h}^{-^*}_{\ell^-,m^-}\dot{h}^+_{\ell^+,m^+}) \\
&= (\alpha_{\ell^-,m^-,\ell^+,m^+} + \beta_{\ell^-,m^-,\ell^+,m^+})\cos(\psi_{\ell^-,m^-,\ell^+,m^+}) \\
&\quad + (\gamma_{\ell^-,m^-,\ell^+,m^+} + \delta_{\ell^-,m^-,\ell^+,m^+})\sin(\psi_{\ell^-,m^-,\ell^+,m^+}).
\end{aligned} \tag{19}$$

Starting from Eq. (14), we can perform an analogous derivation as Ruiz *et al.* [41] demonstrated to get from Eq. (7) to Eq. (10). We obtain the linear momentum flux in $z$ direction in terms of symmetric and antisymmetric amplitudes $a^{\pm}_{\ell,m}$, phases $\phi^{\pm}_{\ell,m}$ and their derivatives,

$$
\begin{aligned}
\frac{\mathrm{d}P_z}{\mathrm{d}t} &= \lim_{r\to\infty}\frac{r^2}{8\pi}\mathrm{Re}\left(\oint \mathrm{d}\Omega l_z \dot{h}^- \dot{h}^{+^*}\right) \\
&= \lim_{r\to\infty}\frac{r^2}{4\pi}\sum_{\ell=2}^{\infty}\sum_{m=0}^{\ell} c_{\ell,m}\mathrm{Re}(\dot{h}^-_{\ell,m}\dot{h}^{+^*}_{\ell,m}) + d_{\ell,m}\mathrm{Re}(\dot{h}^-_{\ell,m}\dot{h}^{+^*}_{\ell-1,m}) + d_{\ell+1,m}\mathrm{Re}(\dot{h}^-_{\ell,m}\dot{h}^{+^*}_{\ell+1,m}) \\
&= \lim_{r\to\infty}\frac{r^2}{4\pi}\sum_{\ell=2}^{\infty}\sum_{m=0}^{\ell}\left(1-\frac{1}{2}\hat{\delta}_{m,0}\right)\{c_{\ell,m}[(\alpha_{\ell,m,\ell,m}+\beta_{\ell,m,\ell,m})\cos(\psi_{\ell,m,\ell,m}) + (\gamma_{\ell,m,\ell,m}+\delta_{\ell,m,\ell,m})\sin(\psi_{\ell,m,\ell,m})] \\
&\quad + d_{\ell,m}[(\alpha_{\ell,m,\ell-1,m}+\beta_{\ell,m,\ell-1,m})\cos(\psi_{\ell,m,\ell-1,m}) + (\gamma_{\ell,m,\ell-1,m}+\delta_{\ell,m,\ell-1,m})\sin(\psi_{\ell,m,\ell-1,m})] \\
&\quad + d_{\ell+1,m}[(\alpha_{\ell,m,\ell+1,m}+\beta_{\ell,m,\ell+1,m})\cos(\psi_{\ell,m,\ell+1,m}) + (\gamma_{\ell,m,\ell+1,m}+\delta_{\ell,m,\ell+1,m})\sin(\psi_{\ell,m,\ell+1,m})]\}. \qquad (20)
\end{aligned}
$$

Note a factor of $1/2$ from the first to the second line, resulting from starting the sum at $m = 0$ by using the symmetries in Eq. (6) and the symmetries of the coefficients defined in Eq. (11). The sum evaluated for $m = 0$ does not have this factor of $1/2$. Therefore, there is a factor with an additional Kronecker delta $\hat{\delta}_{m,0}$ in front of all terms.

Notice two small sources of error when calculating the kick velocity by numerically integrating over Eq. (20). Since we only have a limited waveform time series, a small integration error occurs. However, since the kick is predominantly influenced by the GW emission in the merger-ringdown regime, this can be considered a negligible factor. Moreover, the available list of multipoles is limited. Consequently, the sum must be terminated at a maximum value of $\ell$. Nevertheless, the resulting errors are typically of a relatively minor nature.

Before studying the full equation (20) in more detail in the following sections, we can identify three simple phenomena by including only the dominant $+/-$ waveform. Then Eq. (20) reduces to

$$
\begin{aligned}
\frac{\mathrm{d}P_z}{\mathrm{d}t} = \lim_{r\to\infty}\frac{r^2}{6\pi}[&(\dot{a}^-\dot{a}^+ + a^-a^+\dot{\phi}^-\dot{\phi}^+)\cos(\phi^- - \phi^+) \\
&+ (-a^-\dot{a}^+\dot{\phi}^- + \dot{a}^-a^+\dot{\phi}^+)\sin(\phi^- - \phi^+)], \qquad (21)
\end{aligned}
$$

where we have suppressed the subscript "2, 2" for better readability. We can immediately see that the out-of-plane kick disappears when the antisymmetric waveform is turned off, $a^-(t) = 0$.

In addition, the terms containing amplitude derivatives $\dot{a}^{\pm}$ are small. We can check this by comparing the magnitudes of $a^{\pm}$, $\dot{a}^{\pm}$, and $\dot{\phi}^{\pm}$ in Fig. 5. There we have plotted these quantities using NR waveforms from the SXS:BBH:0976 simulation, a $q = 2$ precessing system. Note that we deliberately omit the units in this plot. Multiplying various combinations according to Eq. (21) together gives the dimensionless linear momentum flux, so that there is no scaling due to the total mass $M$ or the distance from the source $r$. The time array is shifted so that the time of common horizon formation is at $t = 0$.

Given that the terms in front of the sine and $\dot{a}^-\dot{a}^+$ are small because they contain $\dot{a}^{\pm}$, it follows that only the $\beta = a^-a^+\dot{\phi}^-\dot{\phi}^+$ term in front of the cosine is significant. Consequently, the kick is small when the symmetric and antisymmetric waveforms are perpendicular, i.e., $\phi^- - \phi^+ = \pi/2$, and thus the cosine is zero. Analogously, the kick is maximal when the two waveforms are parallel. This is exactly the argument for the wide kick range observed in Fig. 2.

### B. Amplitude and frequency interplay of the $+/-$ waveform

We will now inspect the individual terms of Eq. (20) to show the importance of the interplay of the symmetric and antisymmetric amplitude $a^{\pm}$ and the frequency $\dot{\phi}^{\pm}$, i.e. the dominance of $\beta$.

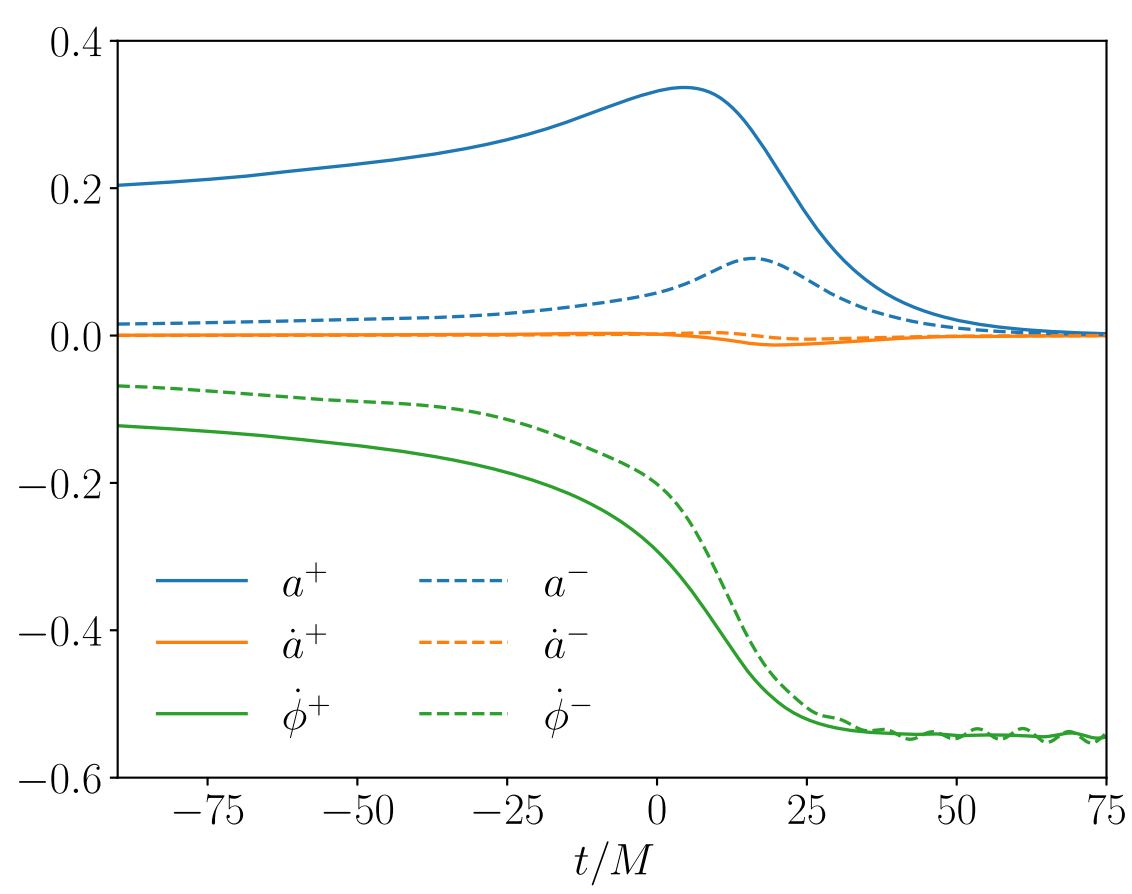

FIG. 5. Dominant symmetric (solid) and antisymmetric (dashed) amplitude and derivative of amplitude and phase in the coprecessing frame. The strongly precessing SXS:BBH:0976 simulation is used. The units on the vertical axis are intentionally suppressed, since in Eq. (21) these functions add up to a dimensionless quantity.

The integral of the $\alpha$, $\beta$, $\delta$, and $\gamma$ terms can be analyzed separately from their corresponding cosine and sine of the phase difference. Then the value of the cosine and sine terms at the time of the $\alpha$, $\beta$, $\delta$, and $\gamma$ maxima serves as the critical factor for estimating the magnitude of the kick.

In order to prove these two statements, we will approximate the integral over Eq. (20), which is equivalent to the kick. The sine and cosine functions are bounded between $-1$ and $+1$. The $\alpha$, $\beta$, $\gamma$, and $\delta$ terms asymptotically go to zero and do not oscillate too strongly. This allows us to write

$$v_z \sim -c_{\ell,m}\Big\{\cos\left[\psi_{\ell,m,\ell,m}(t_{\alpha_{\max}})\right]\int_{-\infty}^{\infty}\mathrm{d}t\alpha_{\ell,m,\ell,m}(t) + \cos\left[\psi_{\ell,m,\ell,m}(t_{\beta_{\max}})\right]\int_{-\infty}^{\infty}\mathrm{d}t\beta_{\ell,m,\ell,m}(t) + \sin\ldots\Big\} + d_{\ell,m}\ldots. \tag{22}$$

Here, we focused on identifying the point in time at which the maximum value of the individual $\alpha$, $\beta$, $\gamma$, and $\delta$ terms was reached,

$$t_{\alpha_{\max}} = \underset{t}{\mathrm{argmax}}[\alpha_{\ell,m,\ell,m}(t)], \tag{23}$$

analogous for the other terms. Note that this is a valid choice since the time range that $\alpha$, $\beta$, $\gamma$, and $\delta$ contributes to the integral is very small.

We have started a systematic study of the individual integrals over $\alpha$, $\beta$, $\gamma$, and $\delta$ using the SXS catalog [49]. We have chosen all precessing systems on quasicircular orbits, where the in-plane spin component for at least one BH is $\chi_{\perp}^{\mathrm{ref},i} \geq 0.3$. It was found that the integral over $\beta_{2,2,2,2}$ dominates the total kick in 97.5% of the simulations. The integral over $\beta_{3,2,2,2}$ and $\beta_{3,3,3,3}$ are likely to be the next leading orders.

As already seen in Sec. IV A, we also found that for the majority of precessing binaries $\dot{a}^{\pm}_{\ell,m}$ is indeed suppressed by orders of magnitude. Since the amplitude derivative does not contribute to $\beta$, this is the reason that the $\beta$ terms dominate significantly over the $\alpha$, $\delta$, and $\gamma$ terms. Another argument for the dominance of $\beta$ is that it is always positive, since the amplitude of a complex quantity and the combination $\dot{\phi}^-\dot{\phi}^+$ is always non-negative.

In Fig. 6 we compare the five most contributing terms, ranked by their integral value, for two strongly precessing binaries. One produces a very low kick of $v_z \approx -14.1$ km/s, `SXS:BBH:0976` (already used in Fig. 5), and one a large kick of $v_z \approx 3025.3$ km/s, `SXS:BBH:1000`. We have transformed the waveform into the coprecessing frame and set the common horizon time to zero.

For these configurations, we immediately see that all terms accumulate during the merger phase at around $t_{\beta_{\max}} \approx 18M$. Also note the power of considering only the dominant waveform quantities, since $\beta_{2,2,2,2}$ is almost an order of magnitude larger than the next leading order contributions. However, both systems have comparable possible achievable kick amplitudes $\beta$, meaning that the interaction between $+/-$ amplitude and $+/-$ frequency is of the same order of magnitude.

### C. Phase difference of the $+/-$ waveform

As previously discussed in Ref. [43], the phase difference between the dominant symmetric and antisymmetric waveform is of significant importance in determining the out-of-plane kick. This finding we confirm to hold true for subdominant multipole asymmetries and for generic precessing systems.

The lower panel of Fig. 6 shows the cosine of the phase differences between different symmetric and antisymmetric multipoles. In the inspiral regime $\cos(\psi_{\ell^-,m^-,\ell^+,m^+})$ oscillates between $-1$ and $+1$. The values that the cosine reaches in the merger-ringdown regime, when the linear momentum flux in $\beta_{\ell^-,m^-,\ell^+,m^+}$ is approximately maximal, is the key to the final kick velocity. For the two chosen simulations it is about $t_{\beta_{\max}} \approx 18M$.

The cosine functions of the low kick `SXS:BBH:0976` simulation evolve in such a way that at $t_{\beta_{\max}} \approx 18M$ they take nearly zero value. An exception is $\cos\left[\psi_{3,2,2,2}(t_{\beta_{\max}})\right]$ and $\cos\left[\psi_{3,3,3,3}(t_{\beta_{\max}})\right]$. However, their signs are reversed. Since their corresponding $\beta$ terms are of the same order, the sum results in an almost vanishing contribution. Note also that this exception is on a very subdominant level compared to the $(\ell^{\pm}, m^{\pm}) = (2,2)$ contribution.

On the other hand, for the very large kick `SXS:BBH:1000` simulation, all phase differences evolve perfectly so that the cosine of them is extremal when the $\beta$ terms are extremal.

Given that the phase difference close to the merger is responsible for the final kick, this quantity in the late ringdown regime is not particularly interesting for the kick. Nevertheless, it is worthwhile to briefly examine this regime. Previous studies [50] noted that the phase difference is locked at merger and shows no further evolution with time in the ringdown phase. This is expected from perturbation theory, since the symmetric and antisymmetric phases should have the same quasinormal frequency. However, recent studies [43] find this to not hold true for all configurations [cf. Fig. 15 on [43] ]. They argue that the Doppler shift to GWs in the ringdown regime due to the kick manifests itself as a linear time dependence in the phase difference. We note similar patterns in Fig. 6, however, the exact relationship between the kick and the phase difference $\psi_{\ell^-,m^-,\ell^+,m^+}$ in the late ringdown regime requires further investigation.

To further study the evolution of the $+/-$ phase differences, a dependence on the in-plane spin angle $\varphi$ needs to be determined. For the dominant phase difference in the PN regime we already have [32,43]

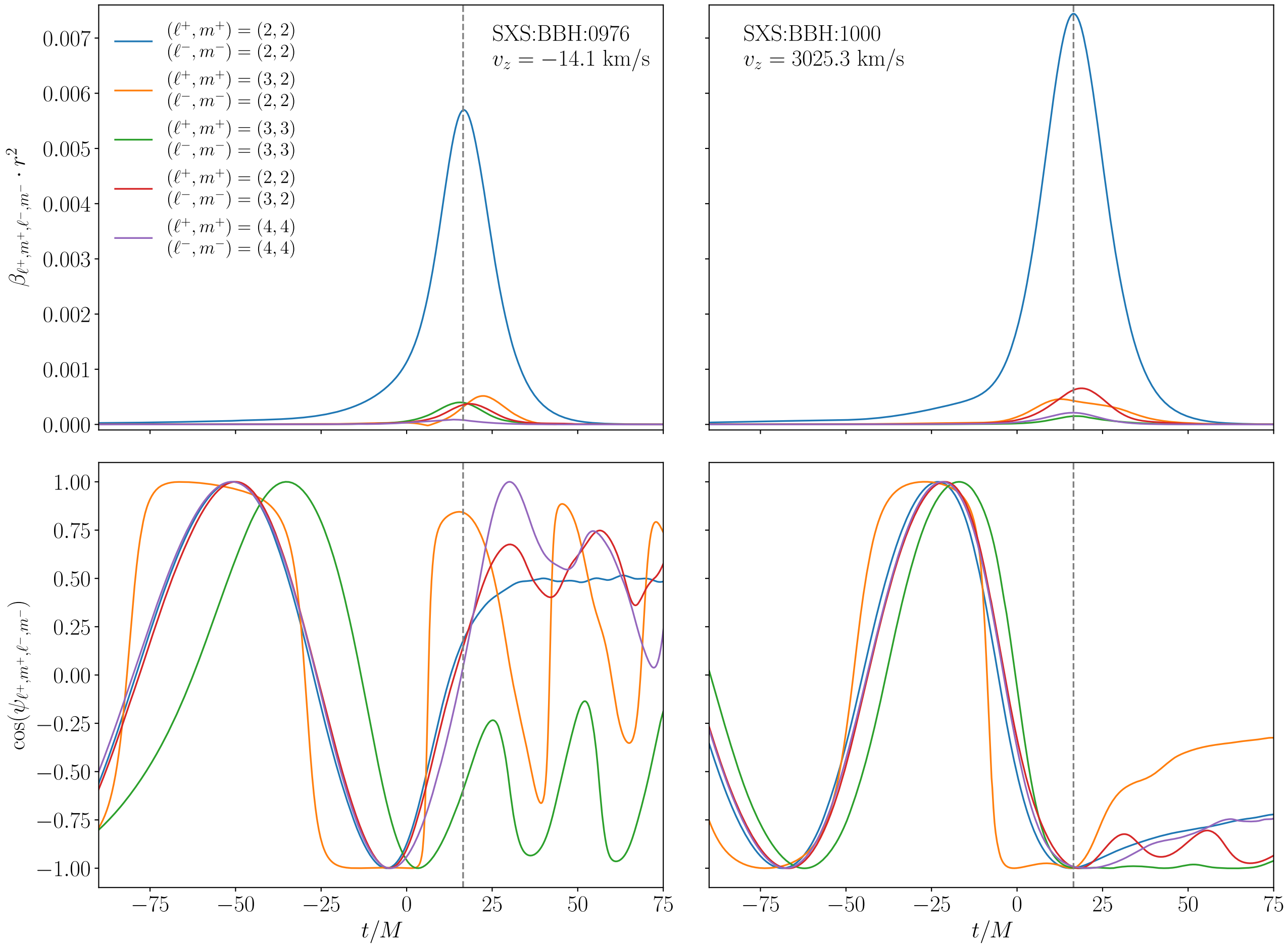

FIG. 6. The most contributing individual terms of Eq. (20) for the SXS:BBH:0976 simulation (left), a highly precessing system with an almost vanishing kick, and for the SXS:BBH:1000 simulation (right), which is the simulation with the largest out-of-plane kick in the SXS catalog. The gray dashed line indicates when the linear momentum flux in $\beta_{\ell^-,m^-,\ell^+,m^+}$ is approximately maximal, at about $t_{\beta_{\max}} \approx 18M$. The value of $\cos(\psi_{\ell^-,m^-,\ell^+,m^+})$ evaluated at $t_{\beta_{\max}}$, weighted with the corresponding possible achievable kick amplitude $\beta$, determines the kick.

$$\psi_{2,2,2,2} = \varphi - \Phi, \tag{24}$$

where $\Phi$ is the orbital phase. In the merger-ringdown regime, which is crucial for the kick, the evolution is not so clear, except that the phase difference is locked to a constant value determined by the exact in-plane spin angle at merger [9]. As already mentioned in Sec. III C a series of future NR simulations with different in-plane spin angles and otherwise fixed intrinsic parameters will also provide more information on the dependence of the individual terms of Eq. (20) on $\varphi$ in the merger-ringdown regime.

## V. CONCLUSIONS

The correlation between kicks and multipole asymmetries has been revisited through the lens of two distinct studies. First, in a phenomenological study, we analyzed a large set of precessing binaries with different spin orientations. We extracted the out-of-plane kick and the antisymmetric waveform using NRSur7dq4 waveforms. Second, in an analytical study, we focused on NR waveforms and compared simulations with large asymmetry but different kicks. Here we summarize our main results:

(i) When plotting the out-of-plane kick against the peak of the amplitude of the dominant multipole asymmetry for a number of precessing binaries, a universal physical characteristic is revealed. The larger the asymmetry is in the merger-ringdown regime, the greater the variation in the achievable kick. This means that large multipole asymmetries do not always induce large kicks. However, in order to generate large kicks, it is necessary to have a large asymmetry combined with optimal spin orientation in the plane such frame-dragging effects are maximal.

(ii) A positive $\chi_{\mathrm{eff}}$ results in a wider out-of-plane kick range compared to a negative $\chi_{\mathrm{eff}}$ for the same

asymmetry. This phenomenon can be related to the orbital hang-up effect.

(iii) We propose to use the spin-asymmetry-kick correlation plots as new performance marker for waveform models that incorporate multipole asymmetries. A study with IMRPhenomXO4a model revealed an issue with the original implementation of the phase of the antisymmetric waveform. With the correct initialization of the reference phase (as detailed in Sec. III C), we found good agreement between IMRPhenomXO4a and the NRSur7dq4 correlation plots.

(iv) We derived a closed-form expression of the linear momentum flux in the $z$ direction in terms of the symmetric and antisymmetric parts of the waveform. The final result is given in Eq. (20) and shows a combination of the symmetric and antisymmetric amplitudes $a^{\pm}_{\ell,m}$, frequencies $\dot{\phi}^{\pm}_{\ell,m}$ and phases $\phi^{\pm}_{\ell,m}$.

(v) The phase difference given by $\psi_{\ell^-,m^-,\ell^+,m^+} = \phi^-_{\ell^-,m^-} - \phi^+_{\ell^+,m^+}$ is the key quantity for the out-of-plane kick of precessing binaries. In particular, the value of $\cos(\psi_{2,2,2,2})$ evaluated at the time when the linear momentum flux peaks determines the direction and magnitude of the kick. In other words, the angle between the symmetric and antisymmetric parts of the dominant $(2,\pm 2)$ multipoles in the merger-ringdown regime determines the kick.

(vi) The maximum possible kick amplitude is mainly determined by the combination of the amplitude of the dominant symmetric and antisymmetric waveforms, expressed in $\beta_{2,2,2,2} = a^-_{2,2} a^+_{2,2} \dot{\phi}^-_{2,2} \dot{\phi}^+_{2,2}$. The antisymmetric amplitude $a^-_{2,2}$ increases with increasing in-plane spin magnitude, while $a^+_{2,2}$ decreases with the mass ratio. Consequently, equal-mass binaries exhibiting a considerable degree of in-plane spin are found to permit a broad range of kicks.

Further research is required to elucidate the relationship between subdominant multipole asymmetries and in-plane spin angles. Subdominant multipole asymmetries appear to exhibit analogous characteristics to those of the dominant contribution to the kick, which has been the primary focus of this analysis. To determine the full out-of-plane kick, these asymmetries must be modeled. To this end, a family of NR simulations with varying in-plane spin angles is essential, as it will enable us to determine the individual kick term dependence on the in-plane spin angle.

Software–NumPy [51], Matplotlib [52], Scipy [53], LALSuite [54], SCRI [44], and GWSurrogate [55].

## ACKNOWLEDGMENTS

We thank Mark Hannam for providing the idea to express the kick in terms of multipole asymmetries. We thank Vijay Varma for pointing out the very informative paper by Ma and his collaborators. This work was supported by an Independent Research Group Grant of the Max Planck Society. Computations were performed on the “Holodeck” cluster of the Max Planck Independent Research Group “Binary Merger Observations and Numerical Relativity”.

## DATA AVAILABILITY

The data supporting this study’s findings are available within the article.